\documentclass[
  ,final            
  ]
  {aipproc}

\layoutstyle{8x11double}

\begin{document}

\title{Eight new MSPs in NGC~6440 and NGC~6441}


\author{Paulo C. C. Freire}{
  address={N.A.I.C., Arecibo Observatory, HC3 Box 53995, PR 00612,
  U.S.A.; {\tt pfreire@naic.edu}}
}

\author{Scott M. Ransom}{
  address={N.R.A.O., 520 Edgemont
  Road, Charlottesville, VA 22903, U.S.A.; {\tt sransom@nrao.edu}}
}

\author{Steve B\'egin}{
  address={Department of Physics and Astronomy, University of
  British Columbia, Vancouver, BC V6T 1Z1, Canada}
}

\author{Ingrid H. Stairs}{
  address={Department of Physics and Astronomy, University of
  British Columbia, Vancouver, BC V6T 1Z1, Canada}
}

\author{Jason W. T. Hessels}{
  address={Astronomical Institute ``Anton Pannekoek'',
  University of Amsterdam, 1098 SJ Amsterdam, The Netherlands}
}

\author{Lucille H. Frey}{
  address={Department of Astronomy, Case Western Reserve
  University, Cleveland, OH 44106}
}

\author{Fernando Camilo}{
  address={Columbia Astrophysics Laboratory, Columbia
  University, New York, NY 10027}
}

\begin{abstract}
We report the discovery of
five new millisecond pulsars in the globular cluster NGC~6440 and
three new ones in NGC~6441; each cluster has one previously known
pulsar. Four of the new pulsars are found in binary systems. One of
the new pulsars, PSR J1748$-$2021B in NGC 6440, is
notable for its eccentric ($e = 0.57$) and wide
($P_b = 20.5$ days) orbit. If the rate of advance of periastron is due
solely to general relativity, we can estimate of the total mass of
this binary system: $2.92 \pm 0.20 M_{\odot}$. This would imply
an anomalously large mass for this pulsar, which could introduce
important constraints in the study of the equation of state for cold
neutron matter.
\end{abstract}


\maketitle


\section{Discovery}

The recent discovery of 30 millisecond pulsars (MSPs) in Terzan 5
\cite{rhs+05,hrs+06}, made using the GBT's S-band receiver and the
pulsar spigot \cite{kel+05} shows that this observing system
(henceforth GBT/S/PS) has an unprecedented sensitivity to MSPs
outside the Arecibo sky. This is only partly due to the factor of 3
higher gain of the GBT as compared to the Parkes telescope.
The higher observing frequency ($\sim$2\,GHz), large observing
bandwidth ($600$\,MHz), and relatively fine frequency resolution
($\sim$0.78\,MHz) provide enough frequency and time resolution to
detect faint MSPs at high dispersion measures (DM). An impressive
demonstration of this is the discovery of PSR~J1748$-$2446ad, the
fastest spinning pulsar known ($P\,=\,1.396\,$ms). This was found in
Terzan~5, where the average DM of the pulsars is 239\,cm$^{-3}$\,pc.
This motivated us to search for MSPs in other promising GCs with this
system.

For the first phase of the survey, we selected the globular clusters
(GCs) NGC~6388, NGC~6440 and NGC~6441 based on their high
stellar interaction rates at the core, $\Gamma_c$ \cite{frb+07}.
Despite the large $\Gamma_c$s of these GCs, only two pulsars were
known in them: PSR~B1745$-$20A (NGC6440A, \cite{lmd96}) and PSR~J1750$-$37A
(NGC~6441A, \cite{pcm+06}). The spin periods and DMs of NGC~6440A and
NGC~6441A are 288.6\,ms and 111.6\,ms and 220 and 234~cm$^{-3}$\,pc,
respectively.  We assumed that the lack of other known pulsars in
these high-DM clusters was largely due to a bias of the earlier
surveys against the detection of MSPs. The GBT/S/PS system can greatly
reduce this bias.

\begin{figure}
  \label{fig:profiles}
  \includegraphics[height=.23\textheight]{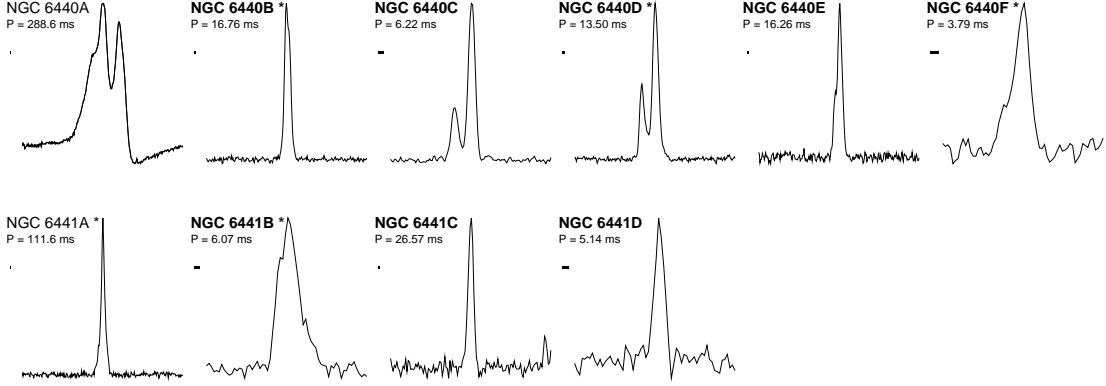}
  \caption{Average 1950 MHz pulse profiles for the 10 pulsars
    known in the GCs NGC~6440 and NGC~6441. These profiles cover one
    full rotation. The pulsars in binary systems have an asterisk
    after their names, and the newly discovered pulsars have their names
    in boldface. The horizontal width of the rectangles
    indicates the system's total time resolution, including the
    effects of dispersive smearing, relative to each pulsar's spin
    period. The dip in power after the main pulse of NGC~6440A might be
    a SPIGOT artifact.}
\end{figure}

With full track observations made in 2005, we discovered eight new
MSPs, five of them in NGC~6440, three in NGC~6441 and none in
NGC~6388. Preliminary results on these objects
were presented in B\'egin (2006)\cite{beg06}. Their pulse profiles,
together with those of the previously known objects, are presented in
Figure~\ref{fig:profiles}. These pulse profiles were obtained by adding
{\em all} detections of the pulsars, carefully excluding all
sub-integrations that are corrupted by RFI.

\begin{figure}[htp]
  \label{fig:fluxes}
  \includegraphics[height=.33\textheight]{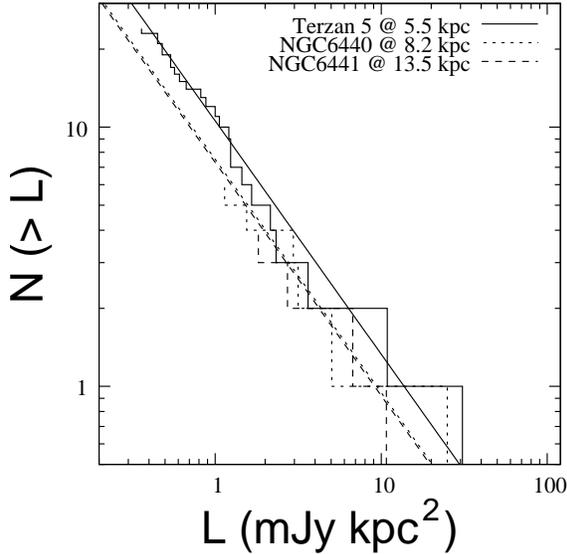}  
  \caption{Cumulative distribution of pseudo luminosities at
    1950 MHz for the pulsars in NGC~6440, NGC~6441 and Terzan 5,
  assuming the most recent distance estimates.}
\end{figure}

\section{Luminosity distribution}

In Figure \ref{fig:fluxes}, we make
cumulative plots of the pseudo luminosities of the pulsars in
NGC~6440, NGC~6441 and Terzan 5, the latter based on the flux density
estimates reported in \cite{rhs+05}. This allows a direct
comparison of the sizes of the pulsar populations of these clusters.
The observing system and observing frequency are the same and the DMs
of the pulsars are very similar,
making the search similarly sensitive to fast-spinning pulsars. The
software used to do the searches is the same, and the software used to
make the flux density measurements is similar (but
not exactly the same). All these factors make the flux
densities directly comparable; the pseudo-luminosities will then be
directly comparable as well if the cluster distances are accurate. We
have used the latest distance estimates for these GCs
(5.5 kpc for Terzan 5, \cite{obb+07}, 8.2 kpc for NGC~6440 and 13.5
kpc for NGC~6441 \cite{vfo07}) to estimate the pulsar
pseudo-luminosities, but caution that these distances still have large
systematic uncertainties.

We have fitted a luminosity function of the type $N(> L) = k_1 L^{-0.9}$
\cite{hrs+07} to the observed pulsar pseudo-luminosity distribution of
each cluster, where $N(> L)$ is the number of pulsars above a pseudo
luminosity of $L$ and $k_1$ is the number of pulsars brighter than 1
mJy kpc$^2$. We obtain $k_1 = 10.6, 7.5$ and $7.3$ for Terzan~5,
NGC~6440 and NGC~6441 respectively. The clusters seem to have
similarly sized pulsar populations, as expected from their stellar
interaction rates: Terzan~5, NGC~6440 and NGC~6441, $\Gamma_c$ have
6.3, 6.3 and 8.6\% of the total stellar interaction rate of
all Galactic GCs. We detect more pulsars in Terzan 5 because that
cluster is closer to us.

\begin{figure}
  \label{fig:mass_mass}
  \includegraphics[height=.45\textheight]{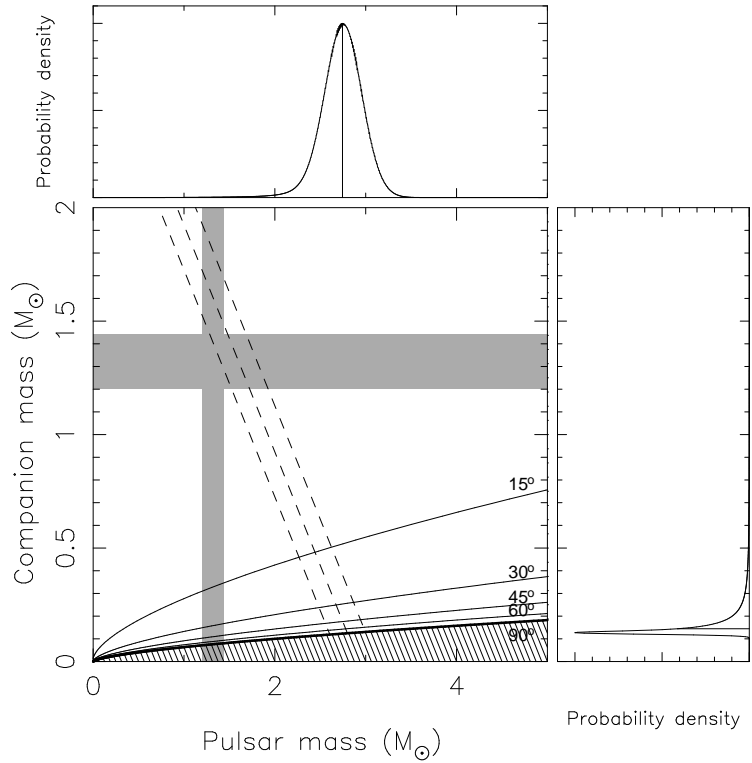}
  \caption{Constraints on the masses of NGC 6440B
  and its companion. The hatched region is excluded by knowledge of
  the mass function and by $\sin i \leq 1$.  The diagonal dashed lines
  correspond to a total system mass that causes a general-relativistic
  $\dot{\omega}$ equal or within 1-$\sigma$ of the measured value.
  The five solid curves indicate constant orbital inclinations.
  The gray bars indicate the range of precisely
  measured neutron star masses, this goes from $\sim$1.20~$M_{\odot}$
  for the companion of PSR~J1756$-$2251\cite{fkl+05} to 1.44~$M_{\odot}$ for
  PSR~B1913+16 \cite{wt03}. We also display the probability density
  function for the mass of the pulsar ({\em top}) and the mass of the
  companion ({\em right}), and mark the respective medians with
  vertical (horizontal) lines.}
\end{figure}

\section{A super-massive neutron star?}

One of the most interesting new discoveries is PSR~J1748$-$2021B. This
is a 16.7-ms pulsar in a wide ($P_b = 20.55\,$days) and
eccentric ($e = 0.57$) orbit with a $\sim 0.1 M_{\odot}$ companion.
This orbital eccentricity has allowed a measurement of the rate of
advance of periastron: $\dot{\omega}\,=\,0.00391(18)^\circ \rm yr^{-1}$.

Assuming that $\dot{\omega}$ is fully relativistic (as argued in
\cite{frb+07}), that implies a total system mass of
$2.92\,\pm\,0.20\,M_{\odot}$. In Figure~\ref{fig:mass_mass},
we present the mass constraints on the components graphically, with
probability distribution functions for the mass of the pulsar and the
companion presented above and on the right respectively \cite{frb+07}.

The total mass of this system suggests that it is a double neutron star
(DNS) binary. However, the probability that it lies within the range of
neutron star masses that have been precisely measured to date,
indicated in Fig.~\ref{fig:mass_mass} by grey bars, is only 0.10\%. This
eventuality would require improbably low orbital inclinations, between $\sim
4^\circ$ and $\sim 5^\circ$.

The median pulsar mass is 2.74~$M_{\odot}$, with the lower and upper
1\,$\sigma$ limits at 2.52 and 2.95~$M_{\odot}$. The pulsar has only
a 1.0\% probability of being less massive than 2.01~$M_{\odot}$, and a
1.0\% probability of being more massive than 3.24~$M_{\odot}$.  If this
high mass value is confirmed, it would be the largest pulsar mass
measured, surpassing that for PSR~B1516+02B in the GC M5
($1.96^{+0.09}_{-0.12} M_{\odot}$, \cite{fwbh07}). Confirmation of
such a large neutron star mass would have profound consequences for
the study of dense matter, since it excludes almost all known
equations of state \cite{lp07}. Since DNS systems have similar masses,
this might even imply that the end
products of the coalescence of DNS systems might themselves be stable
as super-massive neutron stars!


\begin{theacknowledgments}
The National Radio Astronomy Observatory is a facility of the National
Science Foundation operated under cooperative agreement by Associated
Universities, Incorporated.
\end{theacknowledgments}



%

\begin{thebibliography}{10}
\providecommand{\enquote}[1]{``#1''}
\expandafter\ifx\csname url\endcsname\relax
  \def\url#1{\texttt{#1}}\fi
\expandafter\ifx\csname urlprefix\endcsname\relax\def\urlprefix{URL }\fi

\bibitem{rhs+05}
S.~M. {Ransom}, J.~W.~T. {Hessels}, I.~H. {Stairs}, P.~C.~C. {Freire},
  F.~{Camilo}, V.~M. {Kaspi}, and D.~L. {Kaplan}, \emph{Science} \textbf{307},
  892--896 (2005).

\bibitem{hrs+06}
J.~W.~T. {Hessels}, S.~M. {Ransom}, I.~H. {Stairs}, P.~C.~C. {Freire}, V.~M.
  {Kaspi}, and F.~{Camilo}, \emph{Science} \textbf{311}, 1901--1904 (2006).

\bibitem{kel+05}
D.~L. {Kaplan}, R.~P. {Escoffier}, R.~J. {Lacasse}, K.~{O'Neil}, J.~M. {Ford},
  S.~M. {Ransom}, S.~B. {Anderson}, J.~M. {Cordes}, T.~J.~W. {Lazio}, and S.~R.
  {Kulkarni}, \emph{P.A.S.P.} \textbf{117}, 643--653 (2005).

\bibitem{frb+07}
P.~C.~C. {Freire}, S.~M. {Ransom}, S.~{B\'egin}, I.~H. {Stairs},
J.~W.~T.  {Hessels}, L.~H. {Frey}, and F.~{Camilo}, 
  \emph{ArXiv e-prints} \textbf{0711.0925} (2007).

\bibitem{lmd96}
A.~G. Lyne, R.~N. Manchester, and N.~D'Amico, \emph{Ap.J.} \textbf{460}, L41--L44 (1996).

\bibitem{pcm+06}
A.~{Possenti}, A.~{Corongiu}, D.~{Manchester}, F.~{Camilo}, A.~{Lyne},
  N.~{D'Amico}, J.~{Sarkissian}, F.~{Ferraro}, and G.~{Cocozza}, \emph{Chinese
  Journal of Astronomy and Astrophysics Supplement} \textbf{6}, 176--180
  (2006).

\bibitem{beg06}
S.~{B\'egin}, \emph{A Search for Fast Pulsars in Globular Clusters}, Master's
  thesis, University of British Columbia (2006).

\bibitem{obb+07}
S.~{Ortolani}, B.~{Barbuy}, E.~{Bica}, M.~{Zoccali}, and A.~{Renzini},
  \emph{ArXiv e-prints} \textbf{0705.4030} (2007).

\bibitem{vfo07}
E.~{Valenti}, F.~R. {Ferraro}, and L.~{Origlia}, \emph{A.J.} \textbf{133},  1287--1301 (2007).

\bibitem{hrs+07}
J.~W.~T. {Hessels}, S.~M. {Ransom}, I.~H. {Stairs}, V.~M. {Kaspi}, and P.~C.~C.
  {Freire}, \emph{Ap.J.}, \textbf{670}, 363 (2007).

\bibitem{fkl+05}
A.~J. {Faulkner}, M.~{Kramer}, A.~G. {Lyne}, R.~N. {Manchester}, M.~A.
  {McLaughlin}, I.~H. {Stairs}, G.~{Hobbs}, A.~{Possenti}, D.~R. {Lorimer},
  N.~{D'Amico}, F.~{Camilo}, and M.~{Burgay} \emph{Ap.J.Lett.},
  \textbf{618}, L119--L122 (2005).

\bibitem{wt03}
J.~M. {Weisberg}, and J.~H. {Taylor}, \enquote{{The Relativistic Binary Pulsar
  B1913+16},} in \emph{Radio Pulsars}, edited by M.~Bailes, D.~J. Nice, and
  S.~Thorsett, Astronomical Society of the Pacific, San Francisco, 2003, pp.
  93--98.

\bibitem{fwbh07}
P.~C.~C. {Freire}, A.~{Wolszczan}, M.~{van den Berg}, and
J.~W.~T. {Hessels}, \emph{Ap.J.}, submitted (2007).

\bibitem{lp07}
J.~M. {Lattimer}, and M.~{Prakash}, \emph{Phys. Rep.} \textbf{442}, 109--165
  (2007).

\end{thebibliography}
%

\end{document}